\documentclass[modern]{aastex62}

\usepackage{latexsym}		
\usepackage{graphicx}		
\usepackage{rotating}		
\usepackage{natbib}  
\usepackage{savesym}
\usepackage{amssymb}
\usepackage{amsmath}
\usepackage{morefloats}
\savesymbol{doublespace}
\usepackage{xspace}
\usepackage{color}
\usepackage{mdframed}
\usepackage{url}
\usepackage{subfigure}
\usepackage{grffile}
\usepackage{import}
\usepackage[utf8]{inputenc}
\usepackage{booktabs}
\usepackage{fancyhdr}

\usepackage[hang,flushmargin]{footmisc}
\usepackage{ifpdf}

\def\ee#1{\ensuremath{\times10^{#1}}}
\newcommand{\degrees}{\ensuremath{^{\circ}}}

\def\eqref#1{Equation \ref{#1}}

\newenvironment{rotatepage}
{}{}

\begin{document}
\shorttitle{Galactic center star formation \& feedback: key questions}
\shortauthors{Ginsburg et al.}

{\center
\huge
Astro2020 Science White Paper \linebreak

What is the lifecycle of gas and stars in galaxy centers? 
\vspace{5mm}
}

{
\noindent \textbf{Thematic Areas:}

\noindent $\square$ Star and Planet Formation 

\noindent $\square$ Galaxy Evolution  
}

\vspace{15mm}
\correspondingauthor{Adam Ginsburg}\email{adam.g.ginsburg@gmail.com}
\author{Adam Ginsburg}
\author{Elisabeth A.~C.~Mills}
\author{Cara D.~Battersby}
\author{Steven N.~Longmore}
\author{J.~M.~Diederik Kruijssen}


\vspace{-2.0cm}


\section*{Abstract}
\vspace{-1mm}
The closest galaxy center, our own Central Molecular Zone (CMZ; the
central 500 pc of the Milky Way), is a powerful laboratory for studying the
secular processes that shape galaxies across cosmic time, from large-scale gas flows
and star formation to stellar feedback and interaction with a central
supermassive black hole. Research over the last decade has revealed that the
process of converting gas into stars in galaxy centers differs from 
that in galaxy disks. The CMZ is the only galaxy center in which
we can identify and weigh individual forming stars, so it is the key location
to establish the physical laws governing star formation and feedback under the
conditions that dominate star formation across cosmic history. Large-scale surveys
of molecular and atomic gas within the inner kiloparsec of the Milky Way
($\sim10\degrees$) will require efficient mapping capabilities on single-dish
radio telescopes. Characterizing the detailed star formation process will
require large-scale, high-resolution surveys of the protostellar populations
and small-scale gas structure with dedicated surveys on the Atacama Large
Millimeter/submillimeter Array, and eventually with the James Webb Space
Telescope, the Next Generation Very Large Array, and the Origins Space
Telescope.
\vspace{7mm}

\vspace{-10mm}
\begin{figure}[htp]
    \centering
    \includegraphics[width=1.0\textwidth]{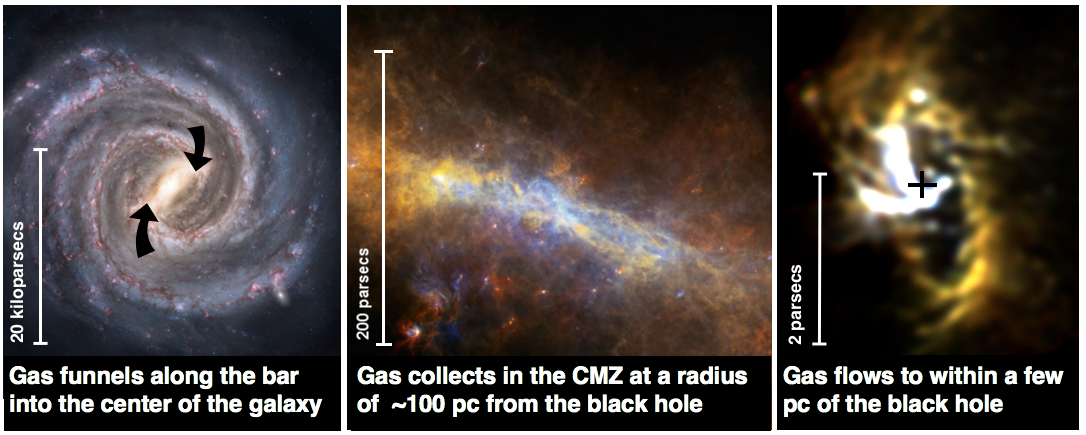}
\end{figure}
\vspace{-10mm}

\clearpage


Our own Central Molecular Zone (CMZ; the inner 500 pc of our Galaxy) is the
closest galaxy nucleus, and it is actively forming stars \citep{Morris1996a}.
It is the only location suitable for studying star formation in hot, turbulent,
high-pressure  \citep{Ao2013a, Ginsburg2016a, Krieger2017a}
molecular gas analogous to the environments of high-redshift galaxies
\citep{Kruijssen2013a}, in which most stars formed. The centers of 
galaxies have more scattered star formation efficiencies than their disks
\citep{Leroy2013a,Usero2015a,Bigiel2016a,Gallagher2018a}, suggesting that
galaxy centers undergo episodic cycles of bursty star formation and quiescence
that may influence the growth of central black holes
\citep[e.g.,][]{Kruijssen2014c,Krumholz2017a,Torrey2017c,Seo2019a}.
We highlight two overarching questions that we can answer over the next decade:
how is our Galactic center fed, and how does it digest gas into stars?

\begin{figure}[htp]
    \includegraphics[width=0.52\textwidth]{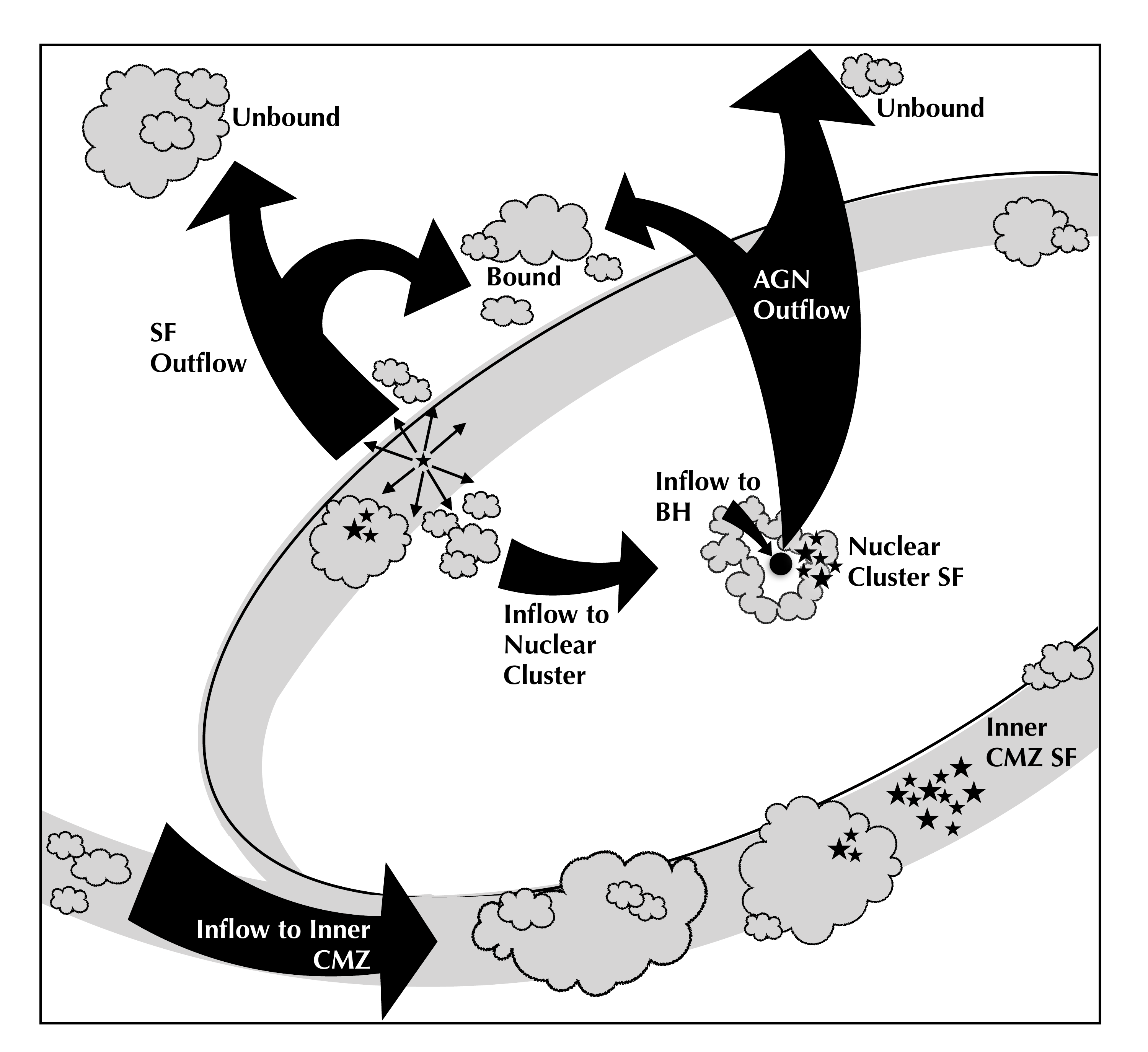}
    \includegraphics[width=0.48\textwidth]{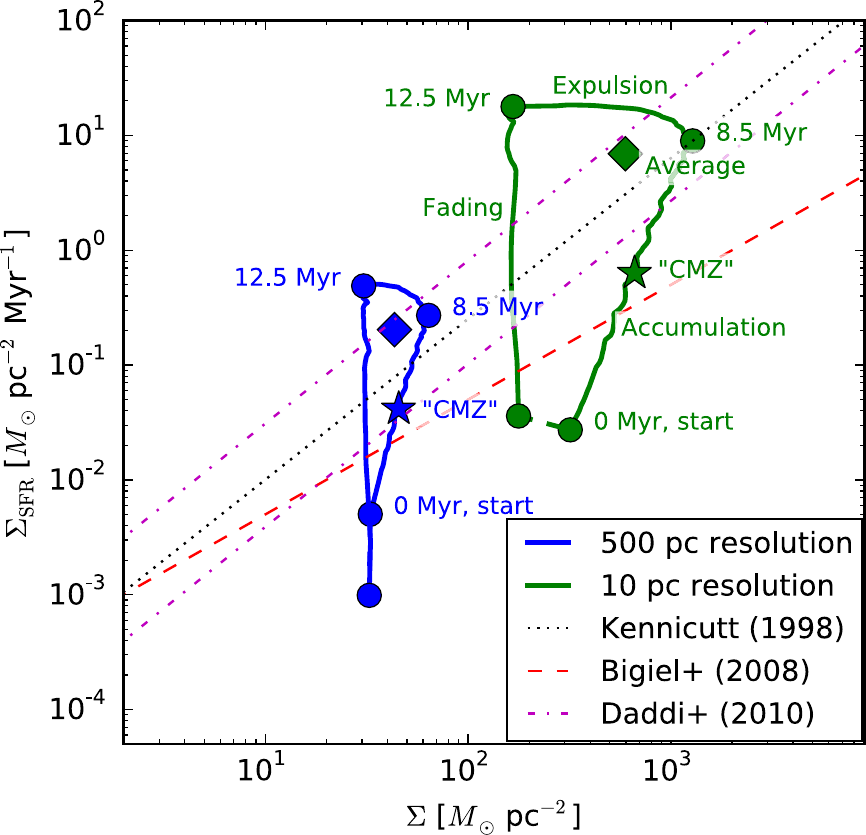}
    \caption{(left) A cartoon of the gas flows in and around the CMZ.
    Inflow from the Galactic bar feeds the CMZ, which contributes
    some material to star formation and some to the central black hole.
    Both the black hole and star formation drive feedback. (right)
    A summary figure from the \citet{Krumholz2015b} dynamical model
    of CMZ gas cycles.  The green and blue curves show the history
    of gas going through accumulation, star formation driven expulsion,
    and fading as the stars go out.  They show these curves in observational
    parameter space of star formation rate surface density as a function of the gas surface density as would be seen at 10 and 500 pc resolution, respectively.
    The dashed curves show the fitted Kennicutt-Schmidt relations
    from \citet{Kennicutt1998a}, \citet{Bigiel2008a}, and \citet{Daddi2010a}.}
\end{figure}

\section{How is gas deposited into the Galactic center from the Galactic disk?}
Fresh material arrives to the Galactic center from the disk via the Galactic bar.
The star formation efficiency in the bar is lower than in much of the disk
\citep[e.g.,][]{Muraoka2019a,Seo2019a}, which means fresh gas can be
transported inward efficiently.
A growing suite of simulations and analytic models
\citep{Krumholz2015b,Sormani2015a,Torrey2017c,Ridley2017a,Krumholz2017a,Sormani2018a,Jeffreson2018b}
demonstrate that galaxy centers vary between brief bursts of star formation 
followed by periods of quiescence given a variety of models of CMZ feeding.
Secular dynamics set a barrier preventing the gas from moving inward,
causing it to build
up at $R_{gal}\sim100$ pc, forming the CMZ \citep{Krumholz2015b}.  Gas
transport further inward, to the nuclear cluster and the central black hole, is
presently small (of the estimated 3\ee{7} solar masses of molecular gas in the
central 500 pc, only 1\ee{4} solar masses is contained within the central $\lesssim5$
pc) and may be driven primarily by stellar feedback
\citep[e.g.,][]{Davies2007a,Kruijssen2017a,Sormani2018b}. The infall through
the final few parsecs is mediated by the circumnuclear disk \citep[CND;
e.g.,][]{Takekawa2017a,Hsieh2017a}, which contains molecular and ionized gas
at temperatures and densities seen nowhere else in the Galaxy \citep{Mills2013b,Mills2017b}.

Key questions about each step of the above inflow process that can be solved in
the next decade include:
\begin{enumerate}
    \item 
\textit{Why is the star formation efficiency low in gas along the bar?}
Very little star formation is currently known in the range $0.1 \mathrm{~kpc} <
R_{gal} < 3 \mathrm{~kpc}$, yet there is abundant molecular gas emission
\citep[e.g.,][]{Dame2001a}.  Apparently, this gas is delivered to the CMZ
with little loss to star formation.  Why is it inefficient
at forming stars until it reaches the Galactic center?  Where is the material
coming from, and how is it deposited on to the CMZ?

    \item 
\textit{How is gas transported from the CMZ to the central few parsecs?}
After gas is delivered into the CMZ, models suggest that it stops flowing inwards, and instead builds up along orbital pathways until it reaches a critical threshold and begins forming stars
\citep{Kruijssen2014c,Krumholz2017a,Sormani2018b,Jeffreson2018b}.  Shortly
afterward, the feedback from high mass stars may dominate over the gas inflow,
resulting in a net loss of gas from the CMZ.  This feedback drives gas both
inward, toward the central black hole, and vertically outward, producing an
outflow. How much of this gas is turned into stars, how much is driven out of
the Galactic center by feedback, and how much is driven towards the central
black hole?  What processes govern each of these steps? 
\end{enumerate}

Our Galactic center is presently at a low state in both star formation and
accretion onto Sgr A*.  The Fermi bubbles suggest there was much higher
activity in the recent past \citep{Su2010a}, but the present-day
outflow activity is low and consistent with driving mostly by star formation
\citep{Law2009a,Law2010a} with some contribution from Sgr A*'s jet
\citep{Muno2008a,Li2013f,Zhu2018b}. Tracking the present-day gas flow from outside
the CMZ to the central black hole is critical
to understand where gas resides during the low-state in galaxies.

\textit{The key advance needed to measure the lifecycle of gas in the Galactic
center and the feeding of the central black hole is an accurate
three-dimensional model of the central $R_{gal}\lesssim100$ pc.} Significant improvements in
our understanding
of the CMZ's geometry have been made in the last decade, refining the structural
model of the molecular gas from a simple ring to a more realistic and physically
self-consistent set of streams
\citep{Molinari2011a,Kruijssen2015a,Ridley2017a,Sormani2018a,Kruijssen2019b}.
Geometric tests are the most important tool to distinguish
between the competing models.

Detailed observational work testing these models has so far progressed only
piecemeal, with case studies assessing the line-of-sight locations of
individual clouds and streams \citep[e.g.,][]{Henshaw2016a,Butterfield2018a}.
Systematic studies
comparing absorption and emission lines at multiple wavelengths, e.g., using
centimeter-wave absorption lines of molecular species like H$_2$CO, OH,
and C$_3$H$_2$, and emission from radio recombination lines, CO, and other bright
molecular transitions in the millimeter range, have the potential to provide
definitive locations of clouds and HII regions and test the theoretical
geometric models.  Assembling a geometric model will require systematic
CMZ-wide (i.e., $\gtrsim2\degrees\times1\degrees$) maps capable of resolving
individual clouds and HII regions ($\lesssim30\arcsec$, $\sim 1\,$pc).  

Much of the molecular gas along the bar has only been observed at extremely
coarse resolution \citep{Dame2001a}. While these observations are useful for
bulk mass inflow measurements \citep[e.g.,][]{Sormani2019a}, they are
inadequate for measurements of turbulence and star formation. Single-dish
molecular line surveys at moderate ($\sim30\arcsec$) resolution covering the
inner $\sim20\degrees$ of the Galaxy are needed to spatially resolve and
measure the turbulent
density and velocity distributions of these clouds. These measurements also
require continued development of the statistical tools to link observed gas
kinematics
and density distributions with the underlying physics
\citep{Koch2017a,Burkhart2018a}.  Observing the gas around the CMZ is essential
for determining how the accretion process, i.e., the interface between bar and
CMZ orbits, affects the observed geometry and controls the inflow process.

\textbf{
Some of the above observational goals can be achieved with existing facilities
and instruments.  However, they require substantial time allocations with
efficient instrumentation (e.g., wide-bandwidth and/or multi-pixel receivers) and
therefore require long-term continued access to single-dish radio telescopes.
}

Multiwavelength observations add important independent information to the above geometric constraints.
The iron K-$\alpha$ lines in the X-ray trace light echoes from individual
high-energy events in the Galactic center, and monitoring and timing these
events can provide direct distance constraints on clouds
\citep[e.g.,][]{Clavel2014a,Churazov2017b,Churazov2017a,Terrier2018a}.
Long-term continued access to X-ray imaging and spectroscopy is needed to
continue these light echo experiments, which grow in power and value over time.
Furthermore, future imaging surveys with the James Webb Space Telescope (JWST) will provide
detailed, high-resolution extinction maps.  Use of JWST in the Galactic center
relies on support for high-dynamic-range imaging, wide-area mapping modes.
Localization of the extincted stars to the Galactic center will enable
molecular cloud mapping analogous to those made of the local kiloparsec with,
e.g., Pan-STARRS \citep{Green2015b}.  Wide-area dust emission and spectral line
mapping with the Origins Space Telescope (OST) will probe the structure of, and turbulence in, CMZ clouds.
Polarization mapping in the sub-mm with OST will be key to understanding the
role of magnetic fields in star formation and gas flows in this region.
By studying the closest example in exquisite detail, we hope to understand the
processes that control the gas flows and star formation in galaxy centers and
the feeding of central supermassive black holes.

\textbf{Multi-wavelength observations of the CMZ require space telescope
observing modes that can cover wide areas with high dynamic range.}

\section{How does star formation change in CMZ-like environments?}
The gas conditions in the Galactic center are substantially different from the
rest of the Galaxy.  Most of the gas mass is in a molecular state
\citep[][]{Kennicutt2012a,Mills2017a} and is warmer and denser than local
molecular clouds, conditions that closely reflect those seen at the peak of the
cosmic star formation history.  The CMZ therefore best represents the dominant
conditions for star formation in the Universe \citep{Kruijssen2013a}.  Star
formation is different under these conditions.  There is continued discussion
that the IMF in the Galactic center massive clusters, the Arches and the
Quintuplet, may be shallower than the canonical IMF \citep{Hosek2019a}. The
molecular gas in the CMZ forms stars at a much lower rate than
comparable-density gas in the Galactic disk \citep{Longmore2013b}, showing that
star formation does not occur at a uniform density threshold
\citep{Kruijssen2014c,Rathborne2014b,Barnes2017b,Walker2018a,Ginsburg2018a}.
Determining what drives the difference between Galactic disk and CMZ
star formation will provide physical insight needed to understand
cosmic star formation.

The key open questions about star formation in Galactic centers are:
\begin{enumerate}
    \item What changes in the star formation process under conditions
        of higher temperature and density (and therefore pressure), 
        cosmic and X-ray backgrounds, and stronger magnetic fields?
        Specifically, how do the star formation rate and the initial
        mass function change under these conditions?
    \item How does episodic star formation progress?  Is star formation
        in galaxy centers tied to specific spatial or temporal triggers,
        and do these triggers represent a distinct mode from galaxy disk star
        formation?
\end{enumerate}

{\it
The key advance needed to address these questions is a complete census of the
ongoing star formation activity in the CMZ, i.e., identification and measurement
of all embedded protostars.}

High-resolution imaging in the millimeter and submillimeter regime across the
entire CMZ is needed to perform a census of the ongoing star formation.  The
CMZ constitutes the most confused and extincted line of sight in the sky, so
the mid-infrared Spitzer survey has been deeply ineffective at measuring star
formation in this region \citep{Koepferl2015a}. Individual programs have
begun to tally the protostars (\citealt{Ginsburg2018a,Lu2019a}; Walker et al.\ in prep.; Barnes et al.\ in
prep.) and the SMA has completed a systematic
survey \citep{Battersby2017b}. However, a systematic deep and high-resolution
survey, such as the proposed ALMA CMZ Exploration Survey (ACES, PI Longmore) is
needed to directly measure the star formation rate via the counting of
protostellar cores. Such a survey will also provide both the gas density and
kinematic measurements needed to characterize the turbulence and link the gas and
stars.

For the coming decade, ALMA and JWST will be the key instruments for providing
the prestellar and stellar census, respectively.  However, because ALMA is limited to the millimeter regime,
it will have limited accuracy when determining embedded stellar properties due to the high dust optical depth in disks; the ngVLA will be required to directly measure
the properties of the most deeply hidden protostars.  Very high-resolution ($\sim100\mathrm{~AU}$)
long-baseline observations with the ngVLA in the 70-120 GHz range will measure
the kinematics of protostellar disks, providing a direct probe of protostellar
masses.  JWST will provide both
a census of the young, recently formed stellar population and detailed dust 
extinction maps.  OST will be needed to obtain luminosity measurements of the
most embedded stars and high-resolution maps of dust density.  The key needs
this science program imposes on the space telescopes are mapping speed
and dynamic range; the Galactic center is full of faint features adjacent
to some of the brightest on the sky, and we need instruments and observing
modes capable of detecting both.

\textbf{Measuring star formation in the CMZ requires high-resolution
observations over wide areas.  ALMA, the JVLA and ngVLA, JWST, and OST provide
these capabilities.  Wide-area, high-resolution, and high-dynamic range capabilities should be 
prioritized.
}

Large-scale mapping of CMZ gas will enable comparison of our Galactic center to
nearby galaxies \citep[e.g.,][]{Leroy2018a}.
Recent and ongoing ALMA
surveys are providing tens of parsec-scale observations of nearby galaxies
covering large areas with a few lines (PHANGS-ALMA, Leroy et al.\ in prep) or
CMZ-scale areas across the entire millimeter spectrum (ALCHEMI).  Analogous
data sets covering our own CMZ are needed to generalize the Galactic work to extragalactic systems.  Surveys of the molecular
chemistry across the CMZ are needed to provide indirect means of understanding
the physical scales that will remain unresolved in external galaxies.
Studies of the CMZ will connect Galactic to extragalactic studies of star
formation, provided that wide-field mapping capabilities are available.

\pagebreak
\bibliographystyle{aasjournal}
\bibliography{extracted}

\end{document}